\documentclass[aps,showpacs,preprint,doublespace,11pt]{revtex4}

\usepackage{amssymb,amsmath,graphicx}
\allowdisplaybreaks
\usepackage{amsfonts}

\newcommand{\mket}[1]{\vert{#1}\rangle}
\newcommand{\mbra}[1]{\langle{#1}\vert}

\begin{document}

\title{Incoherent Noise and Quantum Information Processing%
\vspace{0.5in}}
\author{N. Boulant} \author{J. Emerson} \author{T. F. Havel} \author{D. G. Cory}
\affiliation{Department of Nuclear Engineering, MIT, Cambridge, Massachusetts 02139,
USA}
\author{S. Furuta}
\affiliation{Department of Physics, Cavendish Laboratory, University of Cambridge, UK%
\vspace{0.5in}}

\begin{abstract}
Incoherence in the controlled Hamiltonian is an important limitation
on the precision of coherent control in quantum information processing.
Incoherence can typically be modelled as a distribution of unitary
processes arising from slowly varying experimental
parameters.
We show how it introduces artifacts in quantum process tomography and we explain
how the resulting estimate of the superoperator may not be completely positive.
We then
go on to attack the inverse problem of extracting an effective
distribution of unitaries that characterizes the incoherence via a
perturbation theory analysis of the superoperator eigenvalue spectra.
\end{abstract}


\maketitle

\section{Introduction}
One of the biggest challenges in Quantum Information Processing (QIP) is
the precise control of quantum systems. Errors in the control are
conveniently classified as coherent, decoherent, and incoherent
\cite{PraviaRFI}. Coherent errors are systematic and differ from the
desired operation by a unitary operation. Decoherent errors can be
expressed by completely positive
(CP) superoperators \cite{Alicki} and can be counteracted by techniques
such as Quantum Error Correction (QEC) \cite{Shor,NielsenChuang}.
An incoherent process can also be described by a completely positive
superoperator \cite{PraviaRFI,BoulantEntSwap}.
The apparent non-unitary behavior of
the incoherent process arises due to a distribution over external
experimental parameters. 
The incoherent process is described by a superoperator $S$ acting on Liouville
space which can be written, when acting on columnized density matrices $\mket{\rho}$
obtained by stacking their columns on top of each other from left to right \cite{Havel:03}, as
\begin{eqnarray} 
S&=& \int p(\alpha)\overline{U}(\alpha)\otimes
U(\alpha)d\alpha,\label{Sup}
\end{eqnarray}
where $p(\alpha)$ is a probability density, i.e. the fraction of quantum
systems within an ensemble that sees a given $U(\alpha)$ within an
interval $d\alpha$, $\int p(\alpha)d\alpha = 1$, and $\overline{U}$
denotes the complex conjugate of $U$.  
This decomposition of a CP map into
unitary Kraus operators is sometimes called a random unitary decomposition (RUD)
\cite{LeungThesis}. A RUD exists for an incoherent
process, but such a decomposition is sometimes possible even for a very
general decoherent process \cite{Streater} when there is no correspondence between the unitary
operators in the decomposition and some actual distribution of associated
experimental control parameters $\alpha$. The distinction between the
two therefore is practical, and depends primarily
on the correlation time of the variation of
experimental parameters. If the latter quantity is longer than the inverse
of the typical modulation frequency, the process falls into the class of
incoherent noise \cite{Carr,Hahn}. The point of making this distinction is that, whilst
correcting for decoherent errors requires the full power of QEC, in practice incoherent noise
effects are often reduced directly through the design of the time-dependence of the control fields.
This is possible since the operators underlying the incoherence
$U(\alpha)$ are assumed to be time-independent over the length of the expectation value
measurement. Common approaches for instance in Nuclear Magnetic Resonance (NMR)
include composite and adiabatic pulses \cite{Tycko,Shaka,Levitt,Jones,Pines,Silver}. Furthermore,
the work done by Tycko \cite{Tycko} and Jones \cite{Jones2,Jones3} on composite pulses
finds a great application
in QIP since the schemes proposed are universal and therefore work regarless of the input state.

In this paper, we demonstrate how incoherent errors 
introduce particular limitations to Quantum Process Tomography (QPT)
\cite{NielsenChuang,Childs,BoulantQPT,Cirac} due to the correlations they
introduce with an environment in the
QPT input states. Prior work has been devoted to the study of the
implications of such correlations in the system's reduced dynamics
\cite{Buzek,BuzekErra,Pechukas}.
However, to our knowledge they have not explicitly been studied within the
context of QPT to show that the method may output non-completely
positive (NCP) maps. If the existence or origin of such
noise is unknown, it is shown
in \cite{PraviaRFI} by means of explicit examples how one can eventually
infer qualitative information, say its symmetry, about the probability distribution
underlying the incoherent process from superoperator eigenvalue spectra.
In this paper we tackle the inverse problem of extracting an effective
probability distribution $p(\alpha)$ representing the incoherent
superoperator, given a model for the source of the incoherent noise in the system.
Such information is crucial for counteracting the incoherent errors
\cite{PraviaRFI} which, due to their slow variation would otherwise
persist during the experiment and quickly accumulate.

\section{Incoherent Noise and Quantum Process Tomography}
\label{QPT}
QPT measures the experimental map associated with the implementation of a
desired quantum operation by passing a complete set of input
states through the gate and measuring the corresponding output states (see
Fig.~1). 

\begin{figure}
\centerline{\includegraphics[width=8cm,height=4cm]{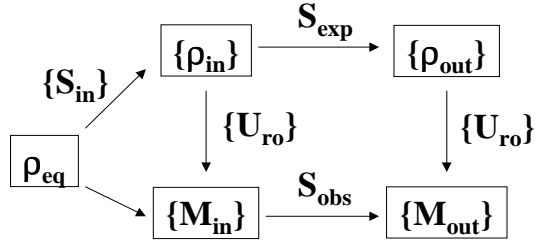}}
\caption{Quantum process tomography method. Starting from thermal equilibrium, a
complete set of input states $\{\rho_{in}\}$ are prepared using a set $\{S_{in}\}$ of
control sequences. The input and output states are measured using a set of readout
pulses $\{U_{ro}\}$ to rotate the density matrix into observable components
$\{M_{in}\}$ and $\{M_{out}\}$. The process to be probed represented by a
superoperator $S_{exp}$ is then applied to these input states to obtain a
corresponding set of output states $\{\rho_{out}\}$. The measured map $S_{obs}$ is
then computed by right multiplying the matrix of output states by the inverted one of
input states (see text for further details).}
\end{figure}

This procedure is important for the experimental study of noise
processes and for the design of quantum error correcting codes
\cite{Shor}. If incoherent noise is present in the preliminary step of QPT,  
then the prepared (input) states will be classically correlated with the
control 
parameter $\alpha$ characterizing the incoherence (Eq 1). Furthermore, if
the correlation time of the noise in the control parameter is long compared to the
coarse-grained time at which the evolution of the system is monitored, then the subsequent
dynamics is non-Markovian \cite{Cohen-Tannoudji}.  More specifically, the gate applied to the
input state (i.e., the gate being characterized by QPT) will be correlated
with the same (slowly-varying) control parameter, and therefore also
correlated with the input state to which it is applied. In such cases,
the measured dynamics are not guaranteed to be completely positive,
and need not correspond to a linear map
\cite{NielsenChuang,Buzek,BuzekErra,Pechukas}.

More generally, any correlations arising from non-Markovian dynamics, 
whether quantum or classical, can lead to incorrect
interpretations of the data obtained from QPT. The ``environment'' which
we assume from the outset is correlated with the system is in general
defined by the degrees of freedom that are not part of the system. For
example, the different spatial locations of individual qubits in a NMR
ensemble, or the bosonic bath producing the fluctuations of the gate
charge in a superconducting qubit \cite{Weiss}.

The basic issues can be seen by exploring the QPT of a spin-$1/2$ system
$A$, coupled to a second spin-$1/2$ system $B$ as its environment.
Borrowing the example used in \cite{Buzek}, the initial density matrix of
the total system may be written as
\begin{eqnarray}
\rho_{AB}=\frac{1}{4}(I_{AB} + \alpha_i \sigma_i\otimes I_B + \beta_j I_A \otimes
\sigma_j + \gamma_{ij} \sigma_i\otimes\sigma_j),
\end{eqnarray}
where $I$ and $\sigma_i$ denote the identity and Pauli matrices respectively. The
density matrix of the system $A$ is then obtained by tracing over the environment $B$:
\begin{eqnarray}
\rho_{A}=\frac{1}{2}(I_{A} + \alpha_i \sigma_i).
\end{eqnarray}
The dynamics of the whole system $\rho_{AB}$ is a unitary evolution $U_{AB}$, so that
the resulting density matrix of $A$ is
\begin{eqnarray*}
\rho'_{A}&=&\sum_{\mu}\mbra{\mu}U_{AB}(\rho_A\otimes\rho_B)U_{AB}^{\dagger}\mket{\mu}\\
& & + \sum_{\mu}\mbra{\mu}U_{AB}\gamma^{'}_{ij}\sigma_i\otimes\sigma_j
U_{AB}^{\dagger}\mket{\mu},
\end{eqnarray*}
where $\gamma^{'}_{ij}=(\gamma_{ij}-\alpha_i\beta_j)/4$ and
$\rho_B=\mathrm{Tr}_A(\rho_{AB})$. Let $\rho_B=\sum_{\nu}p_{\nu}\mket{\nu}\mbra{\nu}$
and $M_{\mu\nu}=\mbra{\mu}U_{AB}\mket{\nu}$. Then the above expression can be
reexpressed as
\begin{eqnarray*}
\rho'_{A}&=&\sum_{\mu\nu}M_{\mu\nu}\rho_AM_{\mu\nu}^{\dagger}\\
& & + \sum_{\mu}\mbra{\mu}U_{AB}\gamma^{'}_{ij}\sigma_i\otimes\sigma_j
U_{AB}^{\dagger}\mket{\mu}
\end{eqnarray*}
The first line therefore corresponds to the Kraus operator sum form
\cite{Kraus2} of
the evolution when initial correlations are not present, while the second
line represents the contribution from these correlations. We can easily
see that there exist $U_{AB}$, e.g., the swap gate, for which these
initial correlations are not observable. In general, however, initial
correlations cause the map to be non-linear or NCP.

Within the context of QPT, we now investigate an explicit example of how
NCP superoperators can arise. We take a set of $4$ initial density
matrices $\rho_{AB}$ such that $\rho_B$ is the same in each case, and
where the input states $\rho_A$ span the Hilbert space of the system $A$
(as required by the QPT procedure)
\begin{eqnarray*}
\rho_{AB}^1&=&(I_{AB} + \beta I\otimes\sigma_z)/4\Rightarrow \rho^1_{A,in}=I/2\\
\rho_{AB}^2&=&(I_{AB} + \alpha \sigma_x\otimes I + \beta I\otimes \sigma_z +
\gamma \sigma_x\otimes\sigma_z)/4\Rightarrow \rho^2_{A,in}=(I+\alpha \sigma_x)/2\\
\rho_{AB}^3&=&(I_{AB} + \alpha \sigma_y\otimes I + \beta I\otimes \sigma_z +
\gamma \sigma_y\otimes\sigma_z)/4\Rightarrow \rho^3_{A,in}=(I+\alpha \sigma_y)/2\\
\rho_{AB}^4&=&(I_{AB} + \alpha \sigma_z\otimes I + \beta I\otimes \sigma_z +
\gamma \sigma_z\otimes\sigma_z)/4\Rightarrow \rho^4_{A,in}=(I+\alpha \sigma_z)/2\\
\end{eqnarray*}
where in each case $\rho_B=(I + \beta \sigma_z)/2$. With the example
$U_{AB}=e^{-i\frac{\pi}{4}\sigma_z\otimes\sigma_z}$,
the corresponding $4$ outputs are $\tilde{\rho}_{A}^1=I/2$, $\tilde{
\rho}_{A}^2=(I+\gamma
\sigma_y)/2$, $\tilde{\rho}_{A}^3=(I-\gamma
\sigma_x)/2$ and $\tilde{\rho}_{A}^4=(I+\alpha\sigma_z)/2$.
We write the density matrices $\rho$ as vectors in Liouville space in the
Zeeman basis
\cite{Havel:03}, which are obtained by first writing the density matrix in the Zeeman
basis and then stacking their columns on top of each other from left to right. We
refer to the resulting vector simply as the ``columnized density matrix'', and will
denote it as a ket $|\rho\rangle$. If we set $\alpha=\beta=0.5$ and $\gamma=0.6$, the
map $S$ is
\begin{equation}
\begin{aligned}[b]
\begin{smallmatrix}  & \textsf{Output 1} &
\textsf{Output 2} & \textsf{Output 3} & \textsf{Output 4}
\end{smallmatrix}\\
\left[ \begin{smallmatrix}
&0.5~~~~&~0.5~~~~~ &~0.5~~~&~0.75 \\[12pt] &0~~~&~0.3i~~~~~&-0.3~~~ &~0~ \\[12pt] &0~~~ &-0.3i~~~~~ &-0.3~~~ &~0~ \\[12pt] &0.5~~ &~0.5~~~~~ &~0.5~~~
&~~0.25
\end{smallmatrix}\right]
\end{aligned}
\begin{aligned}[b]&
\begin{smallmatrix} \hspace{3em} \textsf{Input 1} &
\hspace{0.5em} \textsf{Input 2} & \hspace{0.5em} \textsf{Input 3} & \hspace{0.5em}
\textsf{Input 4}
\end{smallmatrix} \\ &
=S\cdot\left[ \begin{smallmatrix}
&0.5~~~~ &~0.5~~ &~0.5~ &~0.75 \\[12pt] &0~~~ &~0.25~~ &~0.25i~ &~0~ \\[12pt] &0~~~ &~0.25~~ &~-0.25i~ &~0~ \\[12pt] &0.5~~ &~0.5~~ &~0.5~
&~~0.25
\end{smallmatrix} \right]
\end{aligned}
~\leftrightarrow~
\begin{aligned}[b] &
\\S~=~
\left[ \begin{smallmatrix}
&~1 &0 &0 &0 \\[12pt] &0 &~1.2i &0 &0 \\[12pt] &0 &0 &~-1.2i &0 \\[12pt] &0 &0 &0
&~1
\end{smallmatrix} \right]
\end{aligned}\label{supop}
\end{equation}
which is in general non-linear, since one can no longer predict the output
for an arbitrary input state given the action of the gate on these four
specific input states alone.
However, the map can be considered to act linearly on 
the system $A$ Hilbert space, i.e. on the linear combinations of input
states which contain the right correlations with the environment.
For instance, the action of the gate on the input state
$(\rho^2_{A}+\rho^3_{A})/2=(I+\frac{\alpha}{2}(\sigma_x+\sigma_y))/2$ can
be computed by using the above matrix expression if the total input state
is $\rho_{AB}=(I_{AB}+\frac{\alpha}{2}(\sigma_x +
\sigma_y)\otimes I +\beta I\otimes
\sigma_z + \frac{\gamma}{2}(\sigma_x + \sigma_y)\otimes \sigma_z)/4$. This result
conveniently allows one to treat the map as linear. If treated as linear,
the Choi matrix \cite{Choi} ${\mathcal C}=\sum_{i,j=0}^{N-1}(E_{ij}\otimes I)S(I\otimes E_{ij})$, where
$N$ is the dimension of the system's Hilbert space and $E_{ij}$ is the $N\times N$ elementary
matrix (with a "1" in the $ij-th$ position and zeros elsewhere), corresponding to the superoperator
$S$ is not positive semidefinite and consequently $S$ can not be CP
\cite{Havel:03}. 

It is 
suggested in \cite{Havel:03} how the NCP part of the superoperator can be
removed, namely, by removing the negative eigenvalues of the Choi matrix
and then renormalizing so that the trace is equal to the dimension of the
Hilbert space, $N$. We shall call this method CP-filtering. The Choi
matrix corresponding to $S$ in (\ref{supop}) has two non-zero eigenvalues
$(2.2,-0.2)$. In this example, the CP-filtering procedure replaces the
negative eigenvalue by $0$ and then renormalizes the new Choi matrix to
trace $N=2$. The CP-filtering method outputs one unitary Kraus operator.
On the other hand, if no initial correlations were present, the
superoperator would be equal to
$S=\mathrm{diag}(1,0.5i,-0.5i,1)$,
and the corresponding Choi matrix would have two positive eigenvalues
$(0.5,1.5)$, yielding two Kraus operators. This superoperator is therefore
completely positive. Unsurprisingly, we thus observe that the
superoperator obtained from the CP-filtering procedure is not equal to the
superoperator obtained by removing the initial correlations. Therefore,
unless the initial correlations are very small, CP-filtering is a fairly
uncontrolled procedure, giving CP superoperators that may significantly
misrepresent the true quantum dynamics.

We now take the case $\alpha=\beta=\gamma=0.5$ with the same $U_{AB}$ as
previously. Although initial correlations are still present, the
superoperator obtained by the above QPT method, without CP-filtering, is
CP with one Kraus operator. In contrast, if the correlations in the
initial states are removed whilst keeping all other things equal, the
superoperator obtained is CP with two Kraus operators, not one. So even
when initial correlations are present, the superoperator obtained via QPT
may be completely positive. Therefore, one cannot rule out the presence of
initial correlations merely by the existence of a valid Kraus operator sum
form via QPT. Initial correlations can masquerade as CP maps, and in
reality the process may not be linear with respect to arbitrary input
states.

To summarize, the results of this section are: (i) incoherent errors introduce
correlations between the system and the environment in the QPT input states which can persist
during the implemented transformation, (ii) these correlations
can yield non-completely positive
superoperators or non-linear maps, (iii) the CP-filtering method suggested
in \cite{Havel:03} is not equivalent to removing these initial
correlations, and (iiii) initial correlations can masquerade as CP maps
which misrepresent what is in reality a non-linear process with respect to
the input states. See Table
\ref{table} for a summary. This simple analysis explains the
apparent NCP behavior measured in experiments reported in
\cite{BoulantQPT,WeinsteinQFT}. This motivates the need to characterize the incoherent
noise, and to provide ways to correctly interpret QPT data. If the noise
can be successfully characterized, we may use this information to better
counteract the noise in the first place.

\begin{table}
 \begin{tabular}{|c|c|c|c|c|}
 \hline
  \   & CPF & Corr & CP & Num. Kraus Op. \\
  \hline
 Ex 1 & $\times$ & \checkmark & $\times$ & - \\
      & \checkmark    &   \checkmark   & \checkmark &  1 \\
      & $\times$ &    $\times$ &  \checkmark & 2\\
      \hline
 Ex 2 & $\times$ &  \checkmark   & \checkmark & 1 \\
       & $\times$ &  $\times$  & \checkmark & 2 \\
        \hline
 \end{tabular}
 \caption{Summary of the two examples given in section \ref{QPT}.
 CPF is abbreviation for CP-Filtering which is an algorithm for converting a non CP
 superoperator to a CP one. "Corr" is abbreviation for
 correlations (between subsystems $A$ and $B$).
 \label{table}
 }

\end{table}

\section{Extracting Probability Distributions from Superoperator Eigenvalue Spectra}
By applying a first order perturbation theory analysis of the eigenvalues
of superoperators, we now present a method to extract the probability
distribution profile $p(\alpha)$ of unitary matrices present in incoherent
processes. For an incoherent noise to be refocused \cite{Hahn},
knowledge about the noise is \emph{a priori} required. If qualitative
information about the inhomogeneity in the Hamiltonian is known, spectroscopic techniques
can be used to obtain the missing quantitative information. In the
following analysis, we assume that the physical origin of the incoherent
noise is unknown or hidden due to the complexity of the system-apparatus
interactions, but that a mathematical model is presumed.

As presented in the introduction, an incoherent process implies a random
unitary distribution. Incoherent processes are thus unital, which means that the
maximally mixed density matrix is left unchanged. A linear, completely
positive, trace preserving and unital map is called a doubly stochastic
map \cite{Alberti}. Although a single necessary and sufficient condition
for a doubly stochastic map to possess a RUD has not been found to our
knowledge, examples of doubly stochastic maps which do not possess a RUD
are reported in \cite{LeungThesis,Streater}. However, since many
decoherent unital processes can be modelled by a stochastic Hamiltonian,
i.e. semiclassically, we believe that many instances of decoherent
processes can have a RUD.
This belief is supported by the following two facts. Any two density
operators $(\rho,\rho')$ connected by a doubly stochastic map, $\rho'
=\Lambda(\rho)$, can always be related by a transformation of the form
$\rho'=\sum_i p_i U_i\rho U_i^{\dagger}$, where $\sum_i p_i=1$ and $\{U_i\}$
is a set of unitary operators
\cite{Alberti}. Furthermore, all unital maps for a two-level quantum system
always have a RUD \cite{Streater}.

\subsection{Perturbation Theory Analysis of the Eigenvalue Spectra}
In what follows, we take examples from NMR \cite{Ernst} physics where the
main source of incoherence comes from Radio Frequency (RF) power
inhomogeneity. Due to the spatial extent of the sample, individual spins
during the course of a RF field see different powers \cite{PraviaRFI} and
evolve according to different unitary evolutions with different
characteristic frequencies. Note that identical spins can have different
resonance frequencies due to inhomogeneity in the static magnetic field
within the ensemble, which is another source of incoherence. However, as
shown in
\cite{BoulantEntSwap}, the non-unitary features arising from this static
external field inhomogeneity are usually much smaller than those arising
from RF inhomogeneity and will be therefore ignored in this example. 
Finally while the distribution of RF fields can be easily measured via 
a nutation experiment on a single spin, the method presented here is 
quite general and can for instance account for the correlation between
multiple sources of incoherence (several RF fields, DC field etc...).

Let $n$ be the number of spin-1/2 particles in the ensemble, and $U_k$
denote the unitary operator generated by the RF field in the $k$th
frequency interval of the RF amplitude profile. The eigenvalues of the
superoperator $S=\bar U_k
\otimes U_k$ are products of the eigenvalues of $U_k$ with those of $\bar
U_k$. This yields $2^n$ eigenvalues that are equal to unity and $(2^{2n-1}
- 2^{n-1})$ pairs of eigenvalues $(\lambda,\bar{\lambda})$. In general,
the eigenvalues of CP superoperators come in conjugate pairs, but only in
the case of unitary superoperators do all the eigenvalues lie on the
complex unit circle.

The incoherent process resulting from an inhomogeneous distribution of
$U_k$ processes is given by the superoperator $S=\sum_k p_k
\overline{U}_k\otimes U_k$, where $p_k$ is the fraction of spins that sees
the unitary evolution $U_k$. The more broadly the $\{p_k\}$ are
distributed, the larger the degree of inhomogeneity in the evolution, and
the more incoherent noise enters into the evolution. Estimates of the
actual eigenvalues of $S=\sum_k p_k \overline{U}_k\otimes U_k$ will now be
obtained using non-degenerate first-order perturbation theory. Because the
RF pulses are not perfect, even in the absence of RF field inhomogeneity,
we may assume that the unperturbed eigenvalues are generically
non-degenerate. The unitary operator $U_k$ generated by the RF field
acting at position $k$ may be written in exponential form as $U_k = e^{-i
H_k\, t}$ where $H_k$ represents the effective Hamiltonian of the
evolution over the period $t$ for which the pulse is applied ($\hbar$ has been set equal 
to $1$). Defining
$H_0$ to be the unperturbed (and desired) Hamiltonian, the eigenvalues
$\phi_j$ and eigenstates $\mket{\phi_j}$ of $H_{0}$ satisfy the eigenvalue
equation
\begin{equation}
U_{0} \mket{\phi_j}=e^{-i\phi_j\,t}\mket{\phi_j},
\end{equation}
where $U_0 = \exp( -i H_0\, t )$. The Hamiltonian of a particular $U_k$ is assumed to
be a perturbation of the desired, homogeneous Hamiltonian
\begin{equation}
H_k=H_{0} + K_k,
\end{equation}
where $K_k$ is the perturbation. To first order, the new eigenvalues of $H_k$ are
\begin{equation}
\tilde{\phi}_{j,k}=\phi_j+ \mbra{\phi_j} K_k \mket{\phi_j},
\end{equation}
and the corresponding eigenvalues of $U_k$ are
\begin{equation}
U_k\mket{\tilde{\phi}_{j,k}} = e^{-i\tilde{\phi}_{j,k}\,t}\mket{\tilde{\phi}_{j,k}}.
\end{equation}
To first order, the spectral decomposition of
$S$ is
\begin{equation}
S = \sum_{k,m,j} p_k \left( e^{i\tilde{\phi}_{m,k}\,t}
\overline{\mket{\phi_m}}\overline{\mbra{\phi_m}} \otimes
e^{-i\tilde{\phi}_{j,k}\,t}\mket{\phi_j}\mbra{\phi_j}\right)
\end{equation}
and the eigenvalues of $S$ are given approximately by
\begin{eqnarray}
\lambda_{jm}&=&\sum_k p_k e^{-i(\tilde{\phi}_{j,k}-\,\tilde{\phi}_{m,k})\,t}\nonumber\\
\label{eigenvalues}
             &=& e^{-i(\phi_j-\phi_m)t}\sum_k p_k
                                e^{-i\left(\mbra{\phi_j}K_k\mket{\phi_j}-
            \mbra{\phi_m}K_k\mket{\phi_m}\right)t}\,.\nonumber
\end{eqnarray}

We now imagine the scenario where $K_k$ is given by $K_k
t=\frac{\omega_k-\omega_0}{\omega_0}K$. This result would in fact be exact
for one spin on resonance. In this case,
$\frac{\omega_k-\omega_0}{\omega_0}$ is the parameter $\alpha$ defined in
the introduction (which parameterizes the inhomogeneity) and represents
the normalized RF power deviation from the desired power $\omega_0$.
Defining $\Delta\omega = \frac{\omega_k-\omega_0}{\omega_0}$ and
$K_{jm}=\mbra{\phi_j}K\mket{\phi_j}-\mbra{\phi_m}K\mket{\phi_m}$, the
previous equation in the continuous limit becomes
\begin{equation}
\lambda_{jm} = e^{-i(\phi_j-\phi_m)t}\int p(\Delta\omega)
            e^{-iK_{jm}\Delta\omega } d\Delta\omega. \label{data}
\end{equation}
We see in this case that to first order the eigenvalue $\lambda_{jm}$ is just the
unperturbed eigenvalue $e^{-i(\phi_j-\phi_m)t}$ times the Fourier transform of the RF
distribution profile evaluated at $K_{jm}$. This result demonstrates that the
probability distribution profile $p(\Delta\omega)$ of an incoherent process can be
determined, within some degree of approximation, from the eigenvalue structure
$\{\lambda_{jm}\}$ of an experimental superoperator, given some model for the
incoherence $K$. Knowing $K$ would indeed allow one to build the correspondence
between $\lambda_{jm}$ and $K_{jm}$, and then to determine $p(\Delta\omega)$ by
performing an inverse Fourier transform. Of course, this result holds for general $K$
only when the perturbation is in the first order regime, and when the unperturbed
eigenvalues are non-degenerate. But if $K$ approximately commutes with $H_0$, then the
first-order perturbation in the eigenvalues is close to an exact correction, and the
above analysis gives a very accurate description of the incoherent process.

\subsection{Recovery of the Profile}
We now demonstrate via a numerical example how one can recover the profile
$p(\Delta\omega)$ from the eigenvalue spectrum of a measured
superoperator. In the theory derived in the previous subsection,
$\lambda_{jm}$ is the data, i.e., the eigenvalues from the measured
superoperator. A model $K$ is needed for the perturbation, while $\phi_j$
and $\mket{\phi_j}$ are known through the knowledge of $H_0$. Formally
solving for $p(\Delta\omega)$ from (\ref{data}),
\begin{equation}\label{formallysolve}
p(\Delta\omega) = \frac{1}{2\pi} \int \lambda_{jm} e^{i(\phi_j-\phi_m)t}\, e^{i
K_{jm}\Delta\omega} d K_{jm}.
\end{equation}
The different eigenvalues $\lambda_{jm}$ multiplied by $e^{i(\phi_j-\phi_m)t}$
therefore allow us to obtain the complex function of $K_{jm}$ which we shall call $f$,
corresponding to the Fourier transform of the distribution profile.

For the numerical demonstration of this technique we take a $3$-qubit system. We
choose $H_0$ and $K$ such that $|\mbra{\phi_l}K\mket{\phi_n}/(\phi_l-\phi_n)|\simeq
0.1$ for $n\neq l$ and $[H_0,K]\approx 0$ so that first order perturbation theory can
be used \cite{Sakurai}. We then use a measured RF inhomogeneity profile shown in
Fig.~2 to construct the following superoperator acting on Liouville space:
\begin{eqnarray}
S=\sum_{\Delta\omega} p(\Delta\omega) \overline{U}(\Delta\omega)\otimes
U(\Delta\omega).
\end{eqnarray}

\begin{figure}
\centerline{\includegraphics[width=7cm,height=6cm]{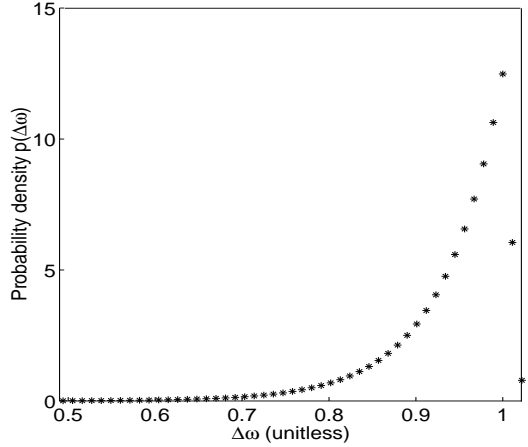}}
\caption{Radio Frequency inhomogeneity profile used to construct the superoperator S
($\int p(\Delta\omega) d\Delta\omega=1$).}
\end{figure}

Provided with this superoperator, we compute its eigenvalues and plot them
on the Argand diagram (see Fig.~3). The perturbation $K$ is such that the
first order limit condition is fulfilled, but that the size of its
diagonal elements in the $H_0$ eigenvectors basis is large enough to
generate significant dephasing and attenuation in the eigenvalues. The
correspondence between $\lambda_{jm}$ and $K_{jm}$ is needed to recover
the distribution profile (\ref{formallysolve}). 
\begin{figure}
\centerline{\includegraphics[width=8cm,height=8cm]{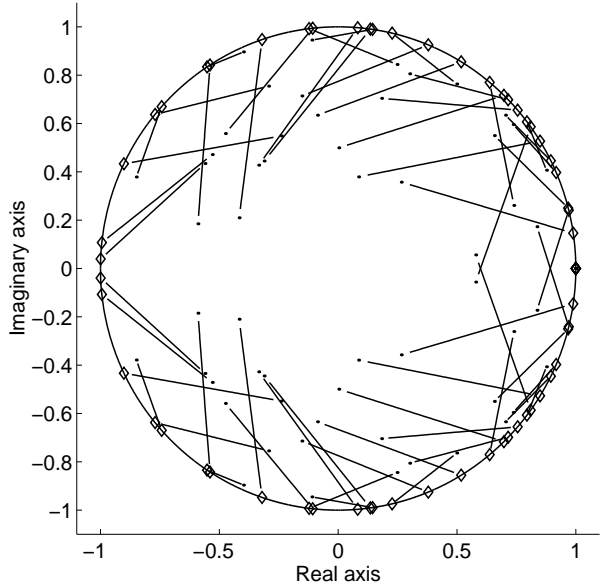}}
\caption{Eigenvalue spectra of the incoherent process $S$ and the desired $S_0$. The
dots are the eigenvalues of the incoherent superoperator $S$ and the diamonds the ones
of the desired unitary superoperator $S_0$ (on the unit circle). Also shown is the
correspondence between the unperturbed and perturbed eigenvalues. The perturbation $K$
is small enough so that the first order limit condition is fulfilled but large enough
to substantially dephase and attenuate the eigenvalues.}
\end{figure}
The correspondence can be
established by first computing
$\mbra{\overline{\phi_j},\phi_m}S\mket{\overline{\phi_j},\phi_m}$, then
searching for the eigenvalue of $S$ closest to it. This allows us to make
the correspondence between one unperturbed eigenvalue with eigenvector
$\mket{\overline{\phi_j},\phi_m}=\mket{\overline{\phi_j}}\otimes\mket{\phi_m}$
(obtained from the knowledge of $H_0$) and one eigenvalue of $S$. The
function $f=\int p(\Delta\omega)e^{-i\Delta\omega K_{jm}}d\Delta\omega$
with respect to $K_{jm}$ can then be constructed. The real and imaginary
parts of that function are plotted in Fig.~4. 
\begin{figure}
\centerline{\includegraphics[width=8cm,height=8cm]{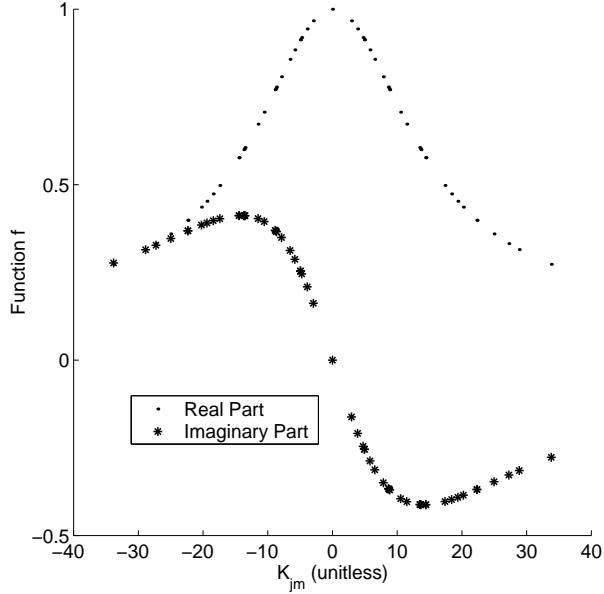}}
\caption{Plot of $f=\int p(\Delta\omega)e^{-i\Delta\omega K_{jm}}$ with respect to
$K_{jm}$ (real and imaginary parts). The function is conjugate symmetric with respect
to $0$ as expected, so that its inverse Fourier transform, which should be a
probability distribution, is real. The point at $K_{jm}=0$ was added to avoid a DC
offset in the reciprocal Fourier domain.}
\end{figure}
Note that we ignore the
degenerate points at $f=1$ because they do not provide any information
about $p(\Delta\omega)$ other than normalization. The $64$ eigenvalues of
the superoperator minus the $8$ degenerate ones equal to $1$ (at
$K_{jm}=0$), plus $1$ eigenvalue added at $K_{jm}=0$ to avoid a DC offset
in the reciprocal domain, yield a complex function $f$ made of $57$
unequally spaced points. The function $f$ is conjugate symmetric with respect to $0$,
which is consistent with the fact that we are supposed to recover a
probability distribution, i.e. a real function, after computing the
inverse Fourier transform. To perform the inverse Fourier transform of a
function sampled at unequally spaced points, we used an algorithm
prescribed in
\cite{MathPaper}. The result is shown in Fig~.5. The width of the
probability distribution and its skewness are recovered to a good extent,
the discrepancy being due to the lack of information about the function
$f$. It is worth mentioning that with $57$ sample points, the window of
$K_{jm}$ values should be large enough to allow low frequency components
of the profile to be reliably extracted. If the incoherent perturbations
were very small, there will not be as many large values of $K_{jm}$, and
therefore less low frequency information would be available. 
\begin{figure}
\centerline{\includegraphics[width=10cm]{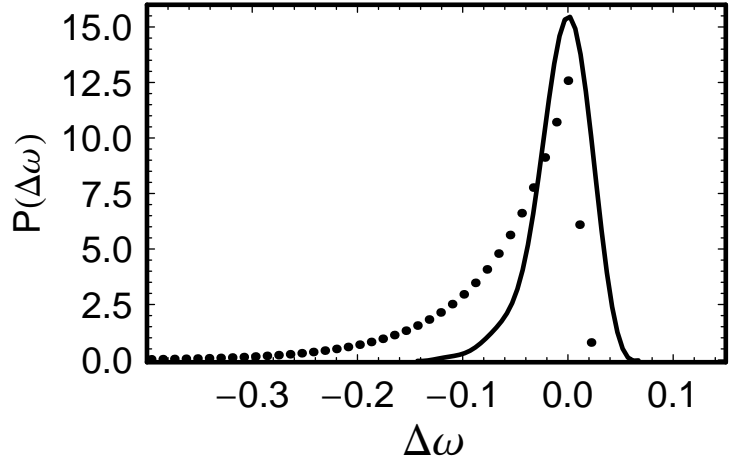}}
\caption{Inverse Fourier transform of the function plotted in Fig.~4 shown together
with points from Fig.~2. The width of the profile in addition to its skewness are
recovered to a good extent.}
\end{figure}
However, the
perturbations can be made larger without changing the mathematical model,
by simply repeating the control sequence several times, provided other
noise mechanisms do not play a significant role. In addition, more points
could be used to get a better sampling resolution, with, for instance, a
$4$-qubit superoperator yielding $241$ points. One cannot have arbitrarily
many points, however, because the correspondence between the eigenvalues
of $S$ and those of $S_0$ could quickly become impossible to establish
unless a very good knowledge of the perturbation $K$ is available. The
density of points in the Argand diagram becomes so large that eigenvalues
can easily become confused. Here, the $3$-qubit superoperator was enough
to recover the essential features of the probability distribution.

If $H_0$ is not exactly known, and in fact a constant offset Hamiltonian
which is proportional to $K$ is present, then a different function $f$ is
obtained :
\begin{equation}
f = \int p(\Delta\omega)e^{-iK_{jm}\Delta\omega }e^{-i\beta K_{jm}}
d\Delta\omega,
\end{equation}
where $\beta$ is a constant real number. Taking the inverse Fourier
transform of $f$ would reveal a distribution $p(\Delta\omega)$ centered
around  $\beta$ rather than $0$, indicating that $H_0+\beta K$ is in fact
the unperturbed Hamiltonian. Perfect knowledge about $H_0$ as a result is
not required provided the offset is approximately proportional to $K$.

It is important for this method to work that the model $K$ chosen \emph{a
priori} is a reasonably faithful one, and that it approximately commutes
with $H_0$. This ensures that the applied first-order perturbation theory
is valid. In the extreme case, in which $K$ anticommutes with $H_0$,
$\mbra{\phi_j}K\mket{\phi_j}=0~\forall j$ and no ``data" would be
available for analysis. It is also worth mentioning that this method is,
needless to say, not scalable. However, in many instances, as in NMR, the
inhomogeneity features are apparatus dependent, so that reasonably
small physical systems can be used to probe them. The scalability of the
method is not necessarily a requirement. The distribution of some control
parameters, once obtained, can be valuable in designing robust
control sequences \cite{PraviaRFI} for larger and more complex
systems.

\section{Conclusion}
Here we reviewed that when incoherence is present during the preparation
of the input states for QPT, the resulting
correlations between the system and the "environment" can play an
important role on the subsequent system's dynamics.
The map obtained by right
multiplying the matrix of output states by the inversion of the matrix of
input states still has a meaning, but a correct interpretation of the
measured data (or transformation) requires an analysis of the incoherence
effects affecting the tomographic procedure. In particular, the measured map
needs not be CP.
If quantitative information is missing, our perturbation
theory analysis of superoperator eigenvalue spectra can be used to
determine an effective distribution of unitaries characterizing the
process, provided a good mathematical model is available. While this
requires a significant effort to measure a 3-qubit superoperator, it is
certainly feasible within present experimental capabilities. Lastly, the
knowledge of the distribution of control parameters should finally allow
us to design more efficient control sequences aimed at counteracting these
deleterious effects.

\section{Acknowledgements}
This work was supported by ARO, DARPA, NSF and the Cambridge-MIT institute.
Correspondence and requests for materials should be addressed to D. G. Cory (e-mail:
dcory@mit.edu). We thank Marcos Saraceno for valuable discussions.


\begin{thebibliography}{11}

\bibitem{PraviaRFI}
M. Pravia, N. Boulant, J. Emerson, E. Fortunato, A. Farid, T. F. Havel, and D. G.
Cory, \textit{J. Chem. Phys.} \textbf{119}, 9993 (2003).
\bibitem{Alicki}
R. Alicki and M. Fannes, \textit{Quantum Dynamical Systems} (Oxford University Press,
Oxford, 2001).
\bibitem{Shor}
P. Shor, \textit{Phys. Rev. A} \textbf{52}, 2493 (1995) ; A. M. Steane, \textit{Phys.
Rev. Lett.} \textbf{77}, 793 (1996) ; J. Preskill, \textit{Proc. Roy. Soc. London A}
\textbf{454}, 385 (1998).
\bibitem{NielsenChuang}
M.~A. Nielsen and I.~L. Chuang, \textit{Quantum Computation and Quantum Information}
(Cambridge University Press, Cambridge, 2001).
\bibitem{BoulantEntSwap}
N. Boulant, K. Edmonds, J. Yang, M. Pravia, and D. G. Cory, \textit{Phys. Rev. A}
\textbf{68}, 032305 (2003).
\bibitem{Havel:03}
T.~F. Havel, \textit{J. Math. Phys.} \textbf{44}, 534 (2003).
\bibitem{LeungThesis}
D. Leung, \textit{Towards Robust Computation}, \textit{Ph.D. thesis}, Stanford
University (2000).
\bibitem{Streater}
L. J. Landau, and R. F. Streater, \textit{Lin. Alg. Appl.} \textbf{193}, 107-127
(1993).
\bibitem{Carr}
H. Y. Carr and E. M. Purcell, \textit{Phys. Rev.} \textbf{93}, 749 (1954).
\bibitem{Hahn}
E. L. Hahn, \textit{Phys. Rev.} \textbf{80}, 580 (1950).
\bibitem{Tycko}
R. Tycko, \textit{Phys. Rev. Lett.} \textbf{51}, 775 (1983).
\bibitem{Shaka}
A. Shaka and R. Freeman, \textit{J. Magn. Reson.} (1969-1992) \textbf{55}, 487 (1983).
\bibitem{Levitt}
M. Levitt, \textit{Prog. Nucl. Magn. Reson. Spectrosc.} \textbf{18}, 61 (1986).
\bibitem{Jones}
H. K. Cummins, G. Llewellyn, and J. Jones, \textit{Phys. Rev. A} \textbf{67}, 042308
(2003).
\bibitem{Pines}
J. Baum, R. Tycko, and A. Pines, \textit{Phys. Rev. A} \textbf{32}, 3435 (1985).
\bibitem{Silver}
M. S. Silver, R. I. Joseph, and D. I. Hoult, \textit{Phys. Rev. A} \textbf{31}, 2753
(1985).
\bibitem{Jones2}
J. A. Jones, quant-ph/0301019.
\bibitem{Jones3}
H. K. Cummins, and J. Jones, \textit{New J. Phys.} \textbf{2}, 6.1-6.12 (2000).
\bibitem{Childs}
A.~M. Childs, I.~L. Chuang, and D.~W. Leung, \textit{Phys. Rev. A} \textbf{64}, 012314
(2001).
\bibitem{BoulantQPT}
N. Boulant, T. F. Havel, M. A. Pravia, and D. G. Cory, \textit{Phys. Rev. A}
\textbf{67}, 042322 (2003).
\bibitem{Cirac}
J. F. Poyatos, J. I. Cirac, and P. Zoller, \textit{Phys. Rev. Lett.} \textbf{78}, 390
(1997).
\bibitem{Buzek}
P. $\check{S}$telmachovi$\check{c}$, and V. Bu$\check{z}$ek, \textit{Phys. Rev. A}
\textbf{64}, 062106 (2001).
\bibitem{BuzekErra}
P. $\check{S}$telmachovi$\check{c}$, and V. Bu$\check{z}$ek, \textit{Phys. Rev. A}
\textbf{67}, 029902 (2003).
\bibitem{Pechukas}
P. Pechukas, \textit{Phys. Rev. Lett.} \textbf{73}, 1060 (1994).
\bibitem{Cohen-Tannoudji}
C. Cohen-Tannoudji, J. Dupont-Roc and G. Grynberg, \textit{Atom-Photon Interactions,
Basic Processes and Applications} (John Wiley \& Sons, New York, 1998).
\bibitem{Weiss}
U. Weiss, \textit{Quantum Dissipative Systems}, Volume 2 of \textit{Series in Modern
Condensed Matter Physics} (World Scientific, Singapore, 1998).
\bibitem{Kraus2}
K. Kraus, \textit{Ann. Phys.} \textbf{64}, 311 (1971).
\bibitem{Choi}
M.~D. Choi, \textit{Lin. Alg. Appl.} \textbf{10}, 285 (1975).
\bibitem{WeinsteinQFT}
Y. Weinstein, T. F. Havel, J. Emerson, N. Boulant, M. Saraceno, S. Lloyd and D. G.
Cory, \textit{in press in J. Chem. Phys.}
\bibitem{Alberti}
P. M. Alberti, and A. Uhlmann, \textit{Stochasticity and Partial Order: Doubly
Stochastic Maps and Unitary Mixing} (Dordrecht, Boston, 1982).
\bibitem{Ernst}
R.~R. Ernst, G. Bodenhausen, and A. Wokaun, \textit{Principles of Nuclear Magnetic
Resonance in One and Two Dimensions} (Oxford University Press, Oxford, 1994).
\bibitem{Sakurai}
J. J. Sakurai, \textit{Modern Quantum Mechanics} (Addison-Wesley, New York, 1994).
\bibitem{MathPaper}
A. Dutt and  V. Rokhlin, \textit{Fast Fourier Transforms for nonequispaced data, Siam
J. Sci. Comput.} \textbf{14(6)} 1368--1393 (1993).


\end{thebibliography}
\end{document}